\documentclass[preprint]{jpsj3}
\usepackage{txfonts}
\usepackage{color}
\usepackage{bm}
\usepackage{cite}

\title{Reduction of the $^{17}$O Knight shift in the Superconducting State and the Heat-up Effect by NMR Pulses on Sr$_2$RuO$_4$}

\author{Kenji Ishida$^1$\thanks{kishida@scphys.kyoto-u.ac.jp}, Masahiro Manago$^{1}$\thanks{present address: Department of Physics, Graduate School of Science, Kobe University}, Katsuki Kinjo$^1$ and Yoshiteru Maeno$^1$}
\inst{$^1$Department of Physics, Graduate School of Science, Kyoto University, Kyoto 606-8502, Japan.} %\\

\abst{
Quite recently, a pronounced drop of $^{17}$O NMR Knight shift in the superconducting (SC) state of an unstrained Sr$_2$RuO$_4$ was reported by Pustogow and Luo ${\it et~al.}$
They revealed such behavior from the free-induction decay (FID) signal after a weak RF pulse. 
We examined this behavior with our single-crystalline Sr$_2$RuO$_4$, and reproduced their result: the peaks of the $^{17}$O-NMR spectra shift in the SC state as long as the RF-pulse power is smaller than a threshold.  
Furthermore, we measured the temperature variation of the Knight shift by a standard spin-echo method with small-power RF pulses, and found that the spin susceptibility decreases in the SC state. 
We conclude that the previous results of the invariance of the Knight shift in the SC state were due to instantaneous destruction of superconductivity by the RF pulses.
The heat-up effect was characterized by the temperature variation of the Knight shift under various measurement conditions. 
}

\begin{document}
\maketitle

\section{Introduction}

The spin-susceptibility measurements by microscopic nuclear magnetic resonance (NMR) Knight shift have played a crucial role for identifying the spin state of the Cooper pair in various superconductors.
This is because the spin susceptibility in the superconducting (SC) state cannot be measured by macroscopic measurements due to the Meissner screening effect.
The reduction of the Knight shift at the Cu and O sites in YBa$_2$Cu$_3$O$_7$ gave a decisive evidence of the spin-singlet pairing\cite{KitaokaJPSJ1988, TakigawaPRB1989, TakigawaPRL1989, BarrettPRB1990}.

Sr$_2$RuO$_4$ is a layered oxide superconductor\cite{MaenoNature1994}, in which the ferromagnetic interaction is considered important in the RuO$_2$ plane\cite{GibbJSSC1974, CaoPRB2003}, and unconventional superconductivity was suggested from various experiments\cite{IshidaPRB1997, MackenziePRL1998}.
Ishida {\it et al.} have performed NMR Knight-shift measurements at the O and Ru sites on Sr$_2$RuO$_4$\cite{IshidaNature1998, MukudaJLTP1999, IshidaPRB2001}, and strongly suggested the spin-triplet paring from the invariant Knight shift across the SC transition temperature $T_c$.
The same conclusion was suggested from the polarized neutron scattering measurement\cite{DuffyPRL2000}. 
In addition, the broken time reversal-symmetry was shown by $\mu$SR and Kerr-effect measurements\cite{LukeNature1998, XiaPRL2006}.
From these experimental results, it has been considered that the spin-triplet chiral $p$-wave pairing is realized in the SC state in Sr$_2$RuO$_4$, which is in good agreement with the theoretical proposal\cite{RiceJPCM1995}. 
It seemed that the pairing state was identified to be the chiral $p$-wave, and that the large stream of the Sr$_2$RuO$_4$ study was shifted to find the novel properties originating from the spin-triplet pairing\cite{MackenzieRMP2003, MaenoJPSJ2012}.  

However, several experiments urging reconsideration of the triplet scenario have also been reported.
The phenomena expected for the chiral $p$-wave state such as the chiral edge current \cite{MatsumotoJPSJ1999, BjornssonPRB2005, KirtleyPRB2007} and splitting of the superconducting transition temperature $T_c$ by in-plane magnetic fields of any magnitude\cite{DeguchiJPSJ2002} have not been detected.
In addition, the first-order (FO) SC-normal (S-N) transition accompanied with a clear hysteresis was observed in a low-temperature region at the upper critical field $H_{c2}$ for fields parallel to the $ab$ plane with the magnetocaloric\cite{YonezawaPRL2013}, specific-heat\cite{YonezawaJPSJ2014} and magnetization measurements\cite{KittakaPRB2014}. 
This abrupt S-N transition suggests a possibility that Sr$_2$RuO$_4$ is a spin-singlet superconductor, because this cannot be interpreted by the conventional orbital depairing effect, but seems to be explained consistently by the Pauli-paramagnetic effect\cite{ClogstonPRL1962, MachidaPRB2008}.
A strong hysteresis and $H_\text{c2}$-suppression were reported in the enhanced superconductivity in the Sr$_2$RuO$_4$-Ru eutectic systems\cite{AndoJPSJ1999} as well as Sr$_2$RuO$_4$ under uniaxial pressure\cite{SteppkeScience2017}.
%The observed $T_c$ maximum occurs at a Lifshitz transition when the Fermi level of the electron $\gamma$ sheet passes through a Van Hove singularity, suggesting that the highly strained sample is in the spin-singlet pairing.

Under such controversial circumstances, Pustogow and Luo {\it et~al.} quite recently reported a pronounced drop of the $^{17}$O NMR Knight shift in the SC state\cite{Pustogow2019}.
They have carried out $^{17}$O-NMR measurements first on a strained sample, and found that the Knight shift decreases but that the magnitude of the Knight-shift decrease depends on the energy of the RF pulses.
Thus, they carefully investigated the dependence of the Knight shift on the energy of single RF pulse by observing the free-induction-decay (FID) signal at 20 mK.
Subsequently, they found that the Knight shift even in the unstrain Sr$_2$RuO$_4$ decreases in the SC state.
We considered that it is crucially important to examine their new results by other groups, and to clarify why the invariant Knight shift was previously reported, if the decrease of the Knight shift below $T_c$ is an intrinsic behavior.
We thus performed $^{17}$O-NMR measurement on the identical  $^{17}$O-enriched single-crystalline Sr$_2$RuO$_4$ to that used in the previous $^{17}$O-NMR measurements\cite{IshidaNature1998, MukudaJLTP1999, ManagoPRB2016}.

\section{Experiment}

Superconducting transition temperature $T_c$ of the $^{17}$O-enriched single-crystalline sample used here is $\sim 1.5$ K. 
This sample was annealed in $^{17}$O 80\% / $^{16}$O 20\% gas in 14 days at 1320 K.       
The sample is a plate shape with the size $\sim 3 \times 10 \times 0.5$ mm$^3$. 
The $c$ axis is perpendicular to the plate; although the $a$ axis is somewhat tilted from the long-axis direction, the field is applied precisely parallel to the $a$ axis of the sample mounted on an adjustable sample holder.
To accommodate the sample in the holder, the size of the NMR coil is $\sim 8 \times 15 \times 2$ mm$^3$ consisting of 55 turns of 0.26-mm-diameter copper wire. 
The NMR coil with the sample was immersed in the $^3$He-$^4$He mixture of the dilution refrigerator (Taiyo-Nissan Product: TS-3H100,  with the cooling power of 20 $\mu$W at 100 mK).
The refrigerator was mounted to the transverse-field SC magnet with a uniaxial goniometer, thus the magnetic field is applied accurately in the $ac$ plane of the sample.

An NMR spectrometer with a 100 W (at 0 dB input) power amplifier (Thamway Product: N146-5049A ) was used for the measurements. 
In general, it is difficult to estimate how much Joule heating occurs after applying an NMR RF pulse, and how much temperature increase occurs instantaneously after the pulse.
This is because various factors such as an impedance of the NMR tank circuit and eddy currents induced by the RF pulses should be taken into consideration; thus in this paper, the energy of the RF pulses related to the Joule heating is expressed by the product of nominal output value of the NMR power amplifier and the pulse duration.
Although the magnitude of the RF-pulse energy absorbed directly by the sample contains ambiguity, the relative ratios of  the absorbed energy to the RF pulse energy would be comparable among different pulse energies.
The effect of the RF-pulse to the superconductivity, particularly whether the superconductivity was destroyed by the RF-pulse energy or not, was investigated by the heat-up test shown in Appendix B.
The instantaneous-temperature increase as well as the recovery to the original temperature is quantitatively discussed in Appendix C. 
We tried to evaluate the eddy current heating on the present sample by using a simple model with the realistic physical parameters of Sr$_2$RuO$_4$. This is discussed in Appendix D.    
 
\begin{figure}[!tbh]
\begin{center}
\includegraphics[width=8.5cm,clip]{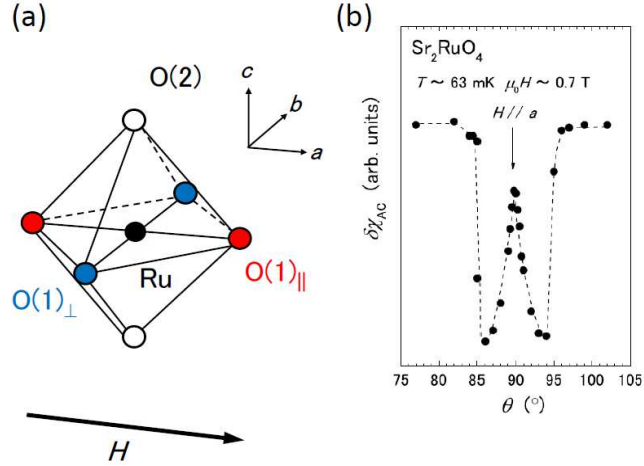}
\end{center}
\caption{(Color online) (a) Three O sites [O(1)$_{\parallel}$, O(1)$_{\perp}$ and O(2) sites] of the RuO$_2$ octahedron in $H \parallel a$. (b) Field-angle dependence of the Meissner screening signal in the $ac$ plane, with the AC field at $f \sim 4.48$ MHz along the $b$ axis. 
$\theta$ is the polar angle between the $c$-axis and DC magnetic-field directions.}
\label{fig1}
\end{figure}

There are two inequivalent O sites in Sr$_2$RuO$_4$, i.e. planer [O(1)] and apical [O(2)] O sites as shown in Fig. \ref{fig1} (a).
When the applied magnetic field $H$ is parallel to the $a$ axis, the O(1) site becomes two distinct sites, O(1)$_{\parallel}$ and O(1)$_{\perp}$, where the $\parallel$ ($\perp$) symbol denotes the O sites with the magnetic field parallel (perpendicular) to the Ru-O-Ru bonds, which is the direction of the principal axis of the maximum electric-field gradient.   
The basal-plane alignment of the sample with respect to the field was done by measuring the field-angle dependence of the Meissner screening signal to be within 0.25$^{\circ}$.
As shown in Fig. \ref{fig1} (b), when the field is parallel to the RuO$_2$ plane, the Meissner signal shows a local minimum due to the vortex-lattice locking in the plane, as reported in a cuprate superconductor\cite{NakaharaiPRB2000}.     
 
%%%%%%%%%%%%%%%%%%%%%%%%%%%%%%%%%%%%%%%%%%%% FIG 1 %%%%%%%%%%%%%%%%%%%%%%%%%%%%
%%%%%%%%%%%%%%%%%%%%%%%%%%%%%%%%%%%%%%%%%%%%%%%%%%%%%%%%%%%%%%%%%%%%%%%%%%%%%%%

\section{Results and discussion}
%%%%%%%%%%%%%%%%%%%%%%%%%%%%%%%%%%%%%%%%%%%% FIG 2 %%%%%%%%%%%%%%%%%%%%%%%%%%%%
\begin{figure}[!tbh]
\begin{center}
\includegraphics[width=8cm,clip]{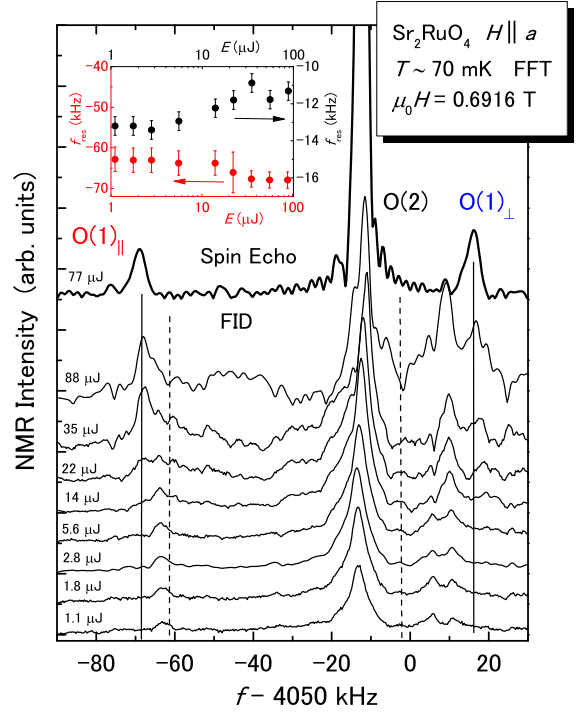}
\end{center}
\caption{(Color online) Free-induction decay (FID) spectra of three central-transition ($I_z : 1/2 \leftrightarrow -1/2$) lines recorded after one RF pulse with various energies. 
In the measurements, the DC external field was applied along the $a$ axis; the pulse width was fixed to be 7 $\mu$s and the pulse intensity was varied. 
The vertical solid lines indicate the peak frequencies in the normal state, and the vertical dotted lines indicate the frequencies at $K$ = 0 of the O(1)$_{\parallel}$ and O(1)$_{\perp}$ sites.
The inset shows the RF-pulse energy dependence of the frequency of the resonance peaks of the O(1)$_{\parallel}$ (red) and O(2) (black) sites at 70 mK.  }
\label{fig2}
\end{figure}
%%%%%%%%%%%%%%%%%%%%%%%%%%%%%%%%%%%%%%%%%%%%%%%%%%%%%%%%%%%%%%%%%%%%%%%%%%%%%%%

Figure \ref{fig2} shows the $^{17}$O-NMR spectra of the central transition of the O(1)$_{\parallel}$, O(2) and O(1)$_{\perp}$ sites in $H$ parallel to the $a$ axis.    
Following Pustogow and Luo {\it et~al.}\cite{Pustogow2019}, the spectra were first measured by the FID method with a fixed pulse duration (7 $\mu$s) and by changing the output of the NMR power amplifier, and compared with the NMR spectrum measured by the spin-echo (SE) method under the same pulse condition (total RF-pulse energy 77 $\mu$J) as the previous measurement.\cite{ManagoPRB2016}
As reported by the reference\cite{Pustogow2019}, the spectral peaks, particularly the O(1)$_{\parallel}$ and O(2) peaks, clearly shift below the threshold energy and stay constant in a smaller energy range as shown in the inset. 
We note that for the O(1)$_{\perp}$site SE spectrum shows a clear single peak, while the FID spectrum shows more complicated peak structure. 
This complication is attributable to the strong coupling between the nuclear and electron spins, which is largest at the O(1)$_{\perp}$; the FID signal in the time domain decayed rapidly and overlapped with the ringing noise just after the RF pulse.
   
The advantages of these FID measurements compared to the SE measurement, conventionally used in the solid-state physics, are that the RF-pulse energy, which makes the sample heat up, can be reduced to less than a half, and that the FID signal can be observed with a smaller-energy RF pulse.
On the other hand, the FID signal is prone to be affected by the ringing noise just after the RF pulse.
In contrast, the advantage of the SE method is that the ringing noise arising from the RF pulses can be removed with the NMR RF-pulse sequence with different-phase pulses.
Nevertheless, the pulse condition for the observation of the SE signal is much severe than that of the FID signal. 

As seen in Fig.~\ref{fig2}, it was difficult to obtain a good shape of the spectrum of the O(1)$_{\perp}$ site by the FID method.
Thus, we tried to record the NMR spectra with the conventional SE measurement, but using long pulses with small voltages to reduce the heat-up effect.
In principle, the $x$ time long $\pi/2$ RF pulse makes the energy of the $\pi/2$-RF pulse reduced to $1/x$, since the $\pi/2$-RF pulse has a relation of $\pi/2 = \gamma_{n}~H_1~t_w \propto \sqrt{P}~t_w$ and $E = P~t_w \propto 1/ t_w$.
Here $\gamma_n$ is the nuclear gyromagnetic ratio, $H_1$ is the oscillating field with the resonance frequency applied perpendicular to the external field, and $t_w$ is the RF-pulse duration satisfying the $\pi/2$ condition.
$P$ and $E$ are the power of the RF pulse, and its energy related to the Joule heating, respectively.        
We performed the heat-up test by these pulses to examine how much reduction in power is needed to assure a negligible effect in the NMR spectrum. 
This test was followed by the idea shown in Fig.~4 in the reference\cite{Pustogow2019}, and examines the time resolved phase change of the low-power phase-detection signal.  
The details of our test-pulse sequence and the results of the test are shown in Appendix B.  
Based on the results of the heat-up test, the SE spectrum was recorded with two 30 $\mu$s-long weak pulses to ensure negligible heating at the SE time position.

%%%%%%%%%%%%%%%%%%%%%%%%%%%%%%%%%%%%%%%%%%%% FIG 3 %%%%%%%%%%%%%%%%%%%%%%%%%%%%
\begin{figure}[!tbh]
\begin{center}
\includegraphics[width=8cm,clip]{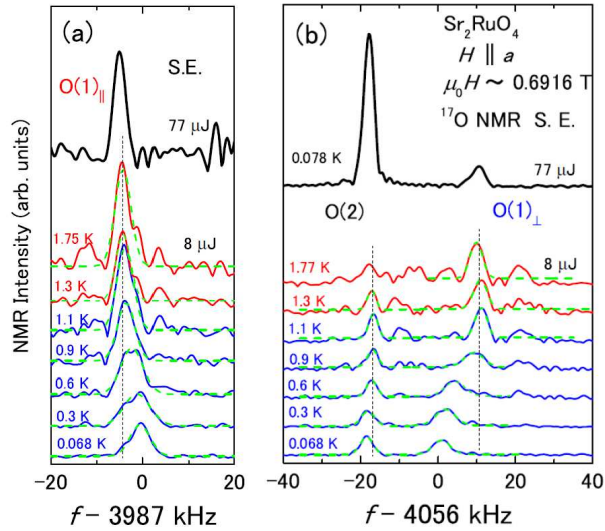}
\end{center}
\caption{(Color online) The temperature variation of the spin-echo (SE) NMR spectra of the (a) O(1)$_{\parallel}$ and (b) O(2) and O(1)$_{\perp}$ sites in $H \parallel a$. These spectra are recorded by SE pulse sequence with two RF pulses with the total energy of 8 $\mu$J.
The SE spectra at 78 mK recorded by two high-energy RF pulses with total power of which is 77 $\mu $J, are also shown.}
\label{fig3}
\end{figure}
%%%%%%%%%%%%%%%%%%%%%%%%%%%%%%%%%%%%%%%%%%%%%%%%%%%%%%%%%%%%%%%%%%%%%%%%%%%%%%%
Figure \ref{fig3} shows the temperature variation of the spin-echo NMR spectra of the (a) O(1)$_{\parallel}$ and (b) O(2) and O(1)$_{\perp}$ sites, respectively, where the spectra are recorded after two RF pulses with the total power of 8 $\mu$J.
They are compared with the spin-echo spectra recorded with the total RF-pulse energy of 77 $\mu$J at 78 mK. 
The NMR spectra recorded with the 8 $\mu$J RF pulses show the systematic dependence on temperature, whereas the peak frequencies for all sites recorded with the total 77 $\mu$J RF pulses at 78 mK remains the same as those above $T_c$. 
This comparison indicates that the electronic state immediately changes into the normal state at the spin-echo time ($\tau \sim 250~\mu$s) after the irradiation of the high-power RF pulses. 
We note that the sample is directly immersed in the liquid $^3$He-$^4$He mixture and the nuclear-spin temperature remains at approximately 78 mK, which we can deduce from the magnitude of the NMR intensity varying inversely proportional to the nuclear spin temperature approximately. 
It is revealed that the previous results of the unchanged Knight shift are ascribed to the instantaneous destruction of the superconductivity by the RF pulses for the NMR observation.
                                  
%%%%%%%%%%%%%%%%%%%%%%%%%%%%%%%%%%%%%%%%%%%% FIG 4 %%%%%%%%%%%%%%%%%%%%%%%%%%%%
\begin{figure}[!tbh]
\begin{center}
\includegraphics[width=8.5cm,clip]{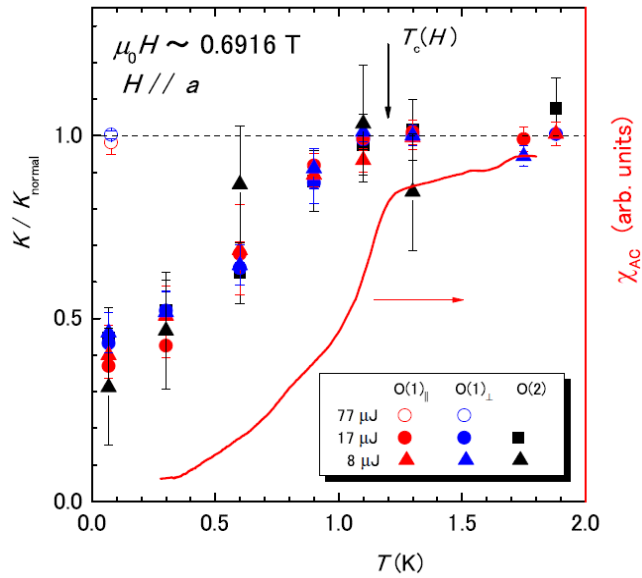}
\end{center}
\caption{(Color online) The temperature dependence of the Knight shift at the O(1)$_{\parallel}$, O(1)$_{\perp}$, and O(2) sites, normalized by the values of the normal-state Knight shift (see text). The temperature variation of the Meissner signal, measured by the variation of the tuning of the NMR tank circuit in the same field, is also shown.}
\label{fig4}
\end{figure}
%%%%%%%%%%%%%%%%%%%%%%%%%%%%%%%%%%%%%%%%%%%%%%%%%%%%%%%%%%%%%%%%%%%%%%%%%%%%%%%
Figure \ref{fig4} shows the temperature dependence of the Knight shift at the O(1)$_{\parallel}$, O(1)$_{\perp}$, and O(2) sites, normalized by the values of the normal-state Knight shifts ($K_{\rm N}$ at O(1)$_{\parallel}$, O(1)$_{\perp}$, and O(2) at $T_c$ is $-0.15$\%, $0.45$\%, and $0.08$\%, respectively).
Each peak measured at various temperatures in Fig.~\ref{fig3} is fitted with the Gaussian function, and the temperature variation of the Knight shift is estimated.           
The temperature dependence of the Knight shift was also measured with the total RF-pulse energy of 17 $\mu$J, and the result is almost same as that measured with the total RF-pulse energy of 8 $\mu$J.
In this figure, the temperature dependence of the Meissner screening signal in the same field is also shown, which was measured by the temperature variation of the tuning frequency of the NMR tank circuit by the network analyzer.
The magnitudes of the Knight shifts of the three O sites decrease at around the onset of the Meissner signal and decreases in the same manner.
Such behavior indicates that the SC diamagnetic effect below $T_c$ is negligibly small,  since the signs of the Knight shifts at the O(1)$_{\parallel}$ and O(1)$_{\perp}$ sites are opposite with each other, while the additional SC diamagnetic effect always gives a negative shift.
The negligibly small SC diamagnetic effect is consistent with the suppression of the Meissner screening signal when the magnetic field is exactly parallel to the $a$ axis as seen in Fig.~\ref{fig1} (b).    
In addition, the orbital Knight shifts at the three sites are also suggested to be small as in the same discussion of the SC diamagnetic effect, where the sign of the orbital Knight shift is positive in general. 
It is noted that the Knight shift measured with the total RF pulse energy of 77 $\mu$J is nearly the same as the value for the normal state due to the instantaneous heat-up.

%%%%%%%%%%%%%%%%%%%%%%%%%%%%%%%%%%%%%%%%%%%% FIG 5 %%%%%%%%%%%%%%%%%%%%%%%%%%%%
\begin{figure}[!tbh]
\begin{center}
\includegraphics[width=8.5cm,clip]{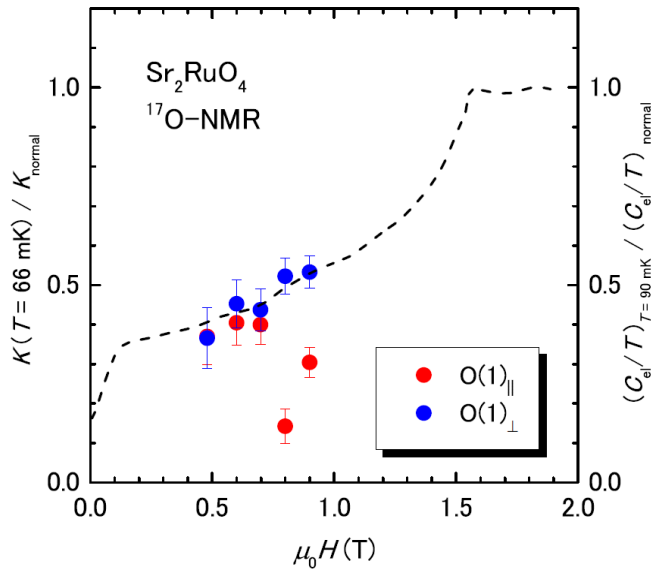}
\end{center}
\caption{(Color online) Field dependence of the residual Knight shift estimated from the comparison between the NMR spectra measured with the total RF-pulse energies of $8 \mu$J and $77 \mu$J, the latter of which is considered to give normal-state Knight shift due to instantaneous heat-up. The field variation of the residual electronic term ($\gamma_{\rm el}$ in the specific heat) is also shown by a dotted curve\cite{NishizakiJPSJ2000}.}
\label{fig5}
\end{figure}
%%%%%%%%%%%%%%%%%%%%%%%%%%%%%%%%%%%%%%%%%%%%%%%%%%%%%%%%%%%%%%%%%%%%%%%%%%%%%%%
By using this heat-up effect, we derived the normal-state Knight shift values, instead of recording the normal-state spectra at the temperatures above $T_c$.
This is useful because it is rather hard to obtain a reliable NMR spectra above $T_c$ with the small-energy RF pulses due to the poor signal to noise ratio above $T_c$ as seen in Fig. \ref{fig3}, since NMR intensity ($I$) follows the Boltzmann relation ($I \times T$= const.). 
This difference in turn indicates that the nuclear temperature is not heated-up so much even after the stronger RF pulses, as we noted above.
Figure \ref{fig5} shows the field dependence of the residual Knight shift estimated from the comparison between the NMR spectra measured with the total RF pulse energy of 8 $\mu$J and those with the high RF pulse energy inducing the instantaneous heat-up effect.  
The field dependence of the residual electronic term $\gamma_{\rm el}$ from the specific heat at $T$ = 0.09 K is overlaid in the figure\cite{NishizakiJPSJ2000}.
It is noted that the overall field dependence of $K / K_{\rm N}$ is similar to that of low-temperature $\gamma_{\rm el}$, although it seems that $K / K_{\rm N}$ at the O(1)$_{\parallel}$ site substantially deviates from this dependences at $\mu_0 H = 0.8$ and $0.9$ T. 
The origin of the deviation has been investigated with further measurements up to $H_{\rm c2}$, which will be presented in a separated paper.
The similar field dependence between the two quantities indicates that the residual $K / K_{\rm N}$ originates from the field-induced quasiparticles such as those associated with vortices.
The semi-quantitative agreement between the ratios of residual $K$ and $\gamma_{\rm el}$ suggests that the spin susceptibility of the Cooper-pair condensate disappears for $H \parallel a$.
However, there are two precautions before concluding the spin symmetry of the Cooper pairs. 
First, the helical spin-triplet states such as $\hat{\boldsymbol{x}}k_x+\hat{\boldsymbol{y}}k_y$ state, giving $K / K_{\rm N} = 0.5$ at $T = 0$ without the Fermi-liquid correction, cannot be excluded completely.
The Fermi-liquid parameter $F_0^a$, which is estimated as $\sim -0.5 $ from the Wilson ratio of $\sim 2$ in Sr$_2$RuO$_4$\cite{SteffensPRL2019}, makes $K / K_{\rm N}$ to reduce from 0.5 to $\sim 0.33$ at $T = 0$ from the relation of $K/K_{\rm N} = [1 + Y(T)](1+F_0^a)/[2 + F_0^a (1 + Y(T)]$, where $Y(T)$ is the Yosida function.
It should be noted that the helical states would exhibit no reduction of $K$ when $H$ is parallel to the $c$ axis.
Second, the spin-singlet state should exhibit the full reduction in $K$ in all field directions, whereas the measurements so far are linked to $H \parallel a$.
Thus, to identify the pairing state thoroughly, the measurement of the Knight shift along the $c$ axis is quite crucial and it is indeed feasible in the strained sample.
This is because the large $H_{c2}$ up to $\sim 1$ T along the $c$ axis in the strained sample, compared with 70 mT of the unstrained sample, enables reliable Knight shift measurements under a large enough field. 
%However, in the case of the helical spin-triplet state, we need other interpretations to understand the experimental results of the $\mu$SR and Kerr-effect measurements, which have been considered as the evidences of the chiral state. 

We comment on why the heat-up effect in Sr$_2$RuO$_4$ is so large that the RF pulse energy of conventional SE measurements destroys the superconductivity. 
First, it is considered that the highly anisotropic vortex state in $H \perp c$, which induces large energy dissipation, would seriously enhance the heat-up effect in the NMR measurement, since the vortices are shaken largely by the RF-pulse field applied perpendicular to the external field and the vortex motion induces the current in superconductors.   
It is also pointed out that Sr$_2$RuO$_4$ is a highly-conductive quasi two-dimensional compound.
The typical inplane residual resistivity is as small as $\rho_{ab} \sim 0.2~\mu \Omega$ cm and the ratio of $\rho_c / \rho_{ab}$ is greater than $10^3$. 
Such anisotropic properties in general make the heating-up effect occurring at the surfaces seriously larger, since the thermal diffusion along the $c$ axis would be small.
In the present Sr$_2$RuO$_4$ case, it is considered that the thermal diffusion in the $c$-direction would be fast enough so that the heating occurs in the entire sample quickly, as discussed in appendix C.  

Finally, we also comment on the results of the Ru-NMR Knight-shift measurements. 
We reported the unchanged or slight-increase behavior of the spin-susceptibility in the SC state from the $^{99}$Ru-NMR Knight-shift measurements\cite{IshidaPRB2001, IshidaPRB2015}.
Comparing between $^{17}$O-NMR and $^{99}$Ru-NMR measurements, it is considered that the heat-up effect in the $^{99}$Ru measurement would be even more serious, because the gyromagnetic ratio of $^{99}$Ru is smaller and the number of $^{99}$Ru nuclei is less than that of $^{17}$O in the present sample, resulting in the larger energy of the RF pulses needed to obtain the signal intense enough to measure the reliable Knight shift.
Nevertheless, since the anomalies related with the SC transition were detected, the $^{99}$Ru-Knight shift data would contain information of the SC state.
To clarify the origin of the observed anomalies, we are planning new $^{99}$Ru-NMR measurements using much smaller-energy RF pulses.            

\section{Conclusion}         

In conclusion, inspired by the recent report by Pustogow and Luo {\it et~al.}, $^{17}$O-NMR Knight shift on Sr$_2$RuO$_4$ was reexamined by the SE method, as well as by the FID method, with the small RF-pulse energy.
Contrary to the previous results, the Knight shift at the crystallographically different two sites decreases below $T_c$ in the same manner. 
Moreover, the magnitude of the residual Knight shift under magnetic fields is comparable to the residual $\gamma_{el}$ in the specific-heat measurements, indicating that the residual Knight shift is mainly ascribed to the field-induced quasi-particles in the SC state. 
These experimental results indicate that the spin susceptibility substantially decreases in the SC state, consistent with the suppression of $H_{c2}$ in the $H - T$ phase diagram and the first order transition near $H_{c2}$.
To identify the paring state thoroughly, measurements of the Knight shift in $H \parallel c$ are needed, which are feasible in the uniaxial strain sample with a higher $T_c$.

\begin{acknowledgment}

%\acknowledgment
The authors deeply appreciate S. Brown for providing them unpublished information.
They would also like to thank Y.~Luo,  A.~P.~Mackenzie, S.~Kitagawa, S.~Yonezawa, H.~Mukuda, Y.~Tokunaga, Y.~Kitaoka, K.~Asayama, Y.~Yanase, Y. Matsuda, and K. Machida for valuable discussions.
This work was supported by Kyoto University LTM Center, and by Grant-in-Aid for Scientific Research (Grant No.~JP15H05745),
Grant-in-Aids for Scientific Research on Innovative Areas ``J-Physics''
(JP15H05882, JP15H05884, and JP15K21732), and ``Topological Materials Science'' (JP15H05851, JP15H0582, JP15K21767) as well as JSPS Core-to-Core program.
M .M. is supported by the Grant-in-Aid for JSPS Research Fellow (Grant No.~19J00336 ) from JSPS.

\end{acknowledgment}

\appendix
\section{The aim of appendices}
In this appendix, we point out a few important points for low-temperature NMR measurements, which we have learned in the present study.
For the low-temperature Knight-shift measurement, the RF-pulse needs to be suppressed not to destroy superconductivity. 
The test of the superconductivity preservation just after the RF pulse can be done by checking the phase of the NMR tank circuit by using an NMR receiver.
This is shown in Appendix B.    
Measurements using the FID signal is effective to reduce the energy of the RF-pulse as long as a clear FID signal is observed.
As described in the text, a longer RF pulse with weaker power is useful for a given tipping angle $\theta$, since the energy of the RF-pulse $E$ is reduced as  $E \propto \theta^2/t_w$, where $t_w$ is the RF-pulse duration time for satisfying $\theta$.
In addition, the sample should have a larger surface area for a larger heat release. 
Since the heat-up occurs instantaneously, a larger surface area might not be very effective for the Knight-shift measurement. 
It is nevertheless very useful for the $T_1$ measurement.
The instantaneous heat up by the RF-pulse and the recovery to the initial temperature is discussed in Appendix C.
We also discuss the eddy-current heating effect on the measured Sr$_2$RuO$_4$ in the normal and SC states in Appendix D.  

\section{Heat-up test}
To examine the effect of heat up by RF pulses for the observation of the SE NMR-signal, we performed the following heat-up test using the same set-up as the NMR measurements.
The test detects a signal closely related to the Meissner screening of the sample at the time when the SE signal is expected to be observed after the irradiation of the same RF pulses for the NMR-signal observation. 
If superconductivity is suppressed by the irradiation of the RF pulses, the Meissner-screening signal would be weakened. 
Such dependence of Meissner-screening signal on the energy of the RF pulses was investigated with the NMR receiver. 
The idea of the test is based on Fig. 4 in the reference\cite{Pustogow2019}.

\begin{figure}[!tbh]
\begin{center}
\includegraphics[width=8.5cm,clip]{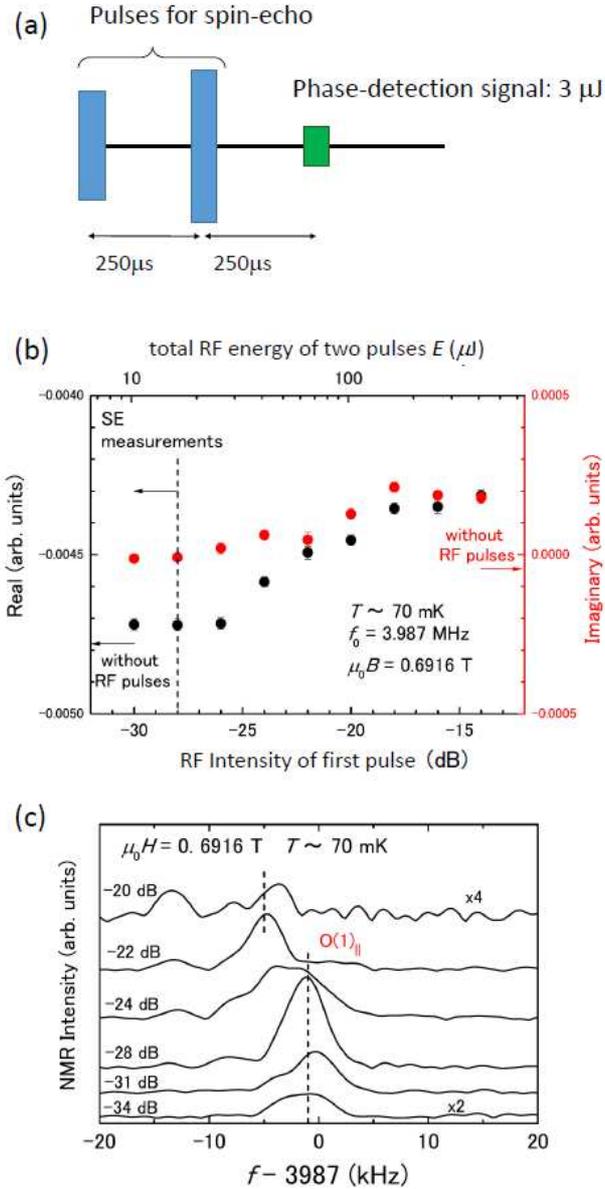}
\end{center}
\caption{(Color online) (a) The pulse sequence for the heat-up test. The heat-up effect by the two pulses for the observation of the NMR spectrum was examined by introducing a phase-detection RF pulse at the spin-echo time position. All three RF pulses have the duration of 30 $\mu$s. The real (in-phase) and imaginary (quadrature) part of the phase-detection RF signal through the NMR tank circuit was measured with the NMR receiver. (b) Dependence of the real (in-phase) and imaginary (quadrature) part of the phase-detection RF signal on the power of the first ($\pi/2$) pulse (bottom axis) and the total energy of two RF pulses (upper axis). (c) RF-power dependence of the spin-echo NMR spectrum of the O(1)$_{\parallel}$ site to examine the heat-up effect by the two RF pulses. 
Here the third pulse was not applied and instead the echo signals from the nuclear spins are Fourier analyzed.
The NMR spectra recorded by the RF pulses smaller than $-28$ dB exhibit the peak at almost the same frequency, but the NMR spectra by the larger energy RF pulses shifts to the lower frequency at which the spectrum at temperature above $T_c$ is observed. }
\label{figA1}
\end{figure}

For our test, the pulse sequence is shown in Fig. \ref{figA1} (a).
As in the standard SE NMR pulse sequence, two pulses separated by the delay time $\tau$ were used; the width of each pulse is fixed to be 30 $\mu$s and the voltage of the second pulse ($\pi$ pulse) is set approximately 1.5 times larger than the first pulse ($\pi$/2 pulse) since the NMR intensity was empirically larger in this pulse condition. 
In the NMR measurements, the spin echo signal appears at time $\tau$ after the second pulse. 
Instead, a weak (3 $\mu$J) phase-detection RF pulse is applied at $\tau$ after the second pulse, and the phase of the weak RF pulse was examined through the NMR tank circuit.
This phase shift is related with the change of the Meissner-screening signal by the irradiation of the RF pulses.  
This procedure was repeated with various total energy of the two NMR RF pulses. 
In this way, information similar to $\chi_{\rm AC}$ at time $\tau$ and at the tank circut frequency is obtained.

Figure \ref{figA1} (b) shows the RF-power dependence of the phase of the detection pulse, where the horizontal bottom axis is relative intensity of the $\pi / 2$ pulse and the horizontal top axis is the total energy of the first two RF pulses.
With increasing RF-pulse intensity from -30 dB (corresponding to the total energy of 10 $\mu$J), the phase was almost constant at the weak-intensity range and started to change at around -26 dB (25 $\mu$J). 

To confirm the validity of this test, we also investigate the RF-intensity dependence of the SE NMR spectrum of the O(1)$_{\parallel}$ site by taking the FFT of the SE signal.
As seen in Fig. \ref{figA1} (c), the peak of the spectrum is almost intensity independent up to $-28$ dB, but shows a double-peak structure at $-24$ dB and shifts to the frequency in the normal state at $-22$ dB.
The spectrum recorded with $-28$dB has a  good shape since the pulse condition is close to the SE condition, but the spectrum with $-20$ dB is weak and distorted due to the large deviation from the proper SE condition.       
Thus, the RF-pulse energy dependence of the phase change of the detection signal is consistent with that of the NMR spectrum.
The test with the SE signal shows the validity of the test with the Meissner-screening signal. 
Based on the results of these tests, we recorded the SE NMR spectrum with the small RF pulses  with the total energy less than 17 $\mu$J corresponding to the $\pi/2$ pulse power of $-28$ dB.    
%%%%%%%%%%%%%%%%%%%%%%%%%%%%%%%%%%%%%%%%%%%% Appendix1%%%%%%%%%%%%%%%%%%%%%%%%%%%%
%%%%%%%%%%%%%%%%%%%%%%%%%%%%%%%%%%%%%%%%%%%%%%%%%%%%%%%%%%%%%%%%%%%%%%%%%%%%%%%

\section{Instantaneous heating up and the recovery to the initial temperature}
It has became apparent that the Knight shift decreases in the SC state of Sr$_2$RuO$_4$.
Since the Knight shift is related to the local spin susceptibility at the nuclear site, which is dominated by the hyperfine couping with the electron spins in the sample, it reflects the temperature of the electronic spin system at the instant when the spectrum is taken.
Using this fact, we estimate the temperature of the electron spin system just after the large-energy RF pulse, and deduce the recovery time of the temperature of the electronic system back to the initial temperature before applying the RF pulse. 
These are done by measuring the time dependence of the NMR spectrum after the RF pulse with the energy large enough to drive the sample into the normal state.

\begin{figure}[!tbh]
\begin{center}
\includegraphics[width=8.5cm,clip]{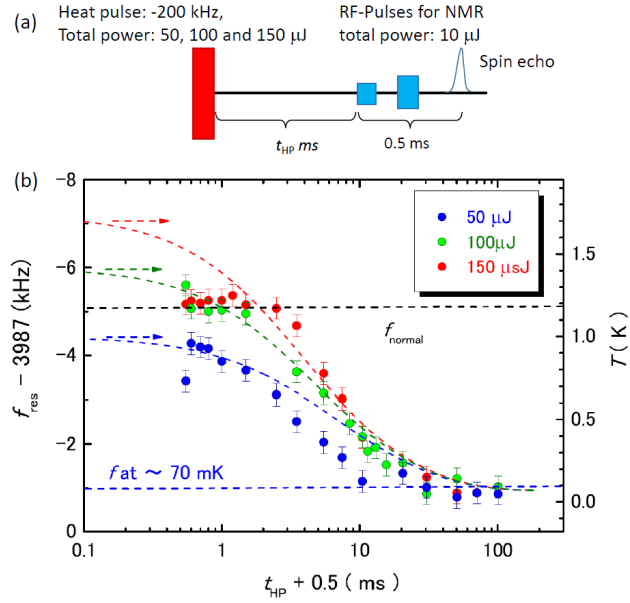}
\end{center}
\caption{(Color online) (a) The pulse sequence to measure the recovery of the electronic-spin temperature after the heat pulse. The frequency of the heat pulse is 3792 kHz at off resonance, so as not to flip the nuclear spins. Time dependence of the resonance frequency was measured by changing the time interval $t_\text{HP}$ between the heat pulse and two NMR pulses. (b) The time $t_\text{HP}$ dependence of the resonance frequency measured after the heat pulse at 50, 100 and 150 $\mu$J. The temperatures after the three heat pulses are simulated based on the simplified model \cite{Walstedt1964}, and are shown by the dotted curves (see the text).    }
\label{figA2}
\end{figure}

The pulse sequence for such measurements is shown in Fig. \ref{figA2} (a).
A large-energy RF pulse, which we call the ``heat pulse'', has a fixed power of $-10$ dB (10 W)  and a variety of energies by changing the pulse duration.
The frequency of the heat pulse is intentional mismatched from the resonance frequency by $-200$ kHz to avoid the nuclear-spin flip by the heat pulse. 
In this way, the heat pulse heats the sample without disturbing the nuclear spin system. 
In order to probe the temperature of the electronic spin system at time $t_{\rm HP}$ after the heat pulse, a SE sequence of RF pulses with small-enough  energy (10 $\mu$ J) was applied then, the NMR spectrum of the O(1)$_{\parallel}$ site was obtained from the Fourier transform of the echo signal.
The NMR resonance frequency obtained in this way with varying $t_\text{HP}$ reflects the temperature evolution from the normal state to the initial temperature, since the Knight shift, i.e. the resonance frequency is in the roughly linear relation to temperature below $T_{c}(H)$ as seen in Fig.~\ref{fig4}.
To characterize the relaxation of the electronic spin temperature, the time $t_{\rm HP}$ is varied from $50~\mu$s to 100 ms.

It is revealed that the temperature of the electronic-spin system immediately increases to above $T_c(H)$ after the heat pulse of 150 or 100 $\mu$J, since the resonance frequency of the O(1)$_{\parallel}$ NMR peak after the heat pulse was observed at the frequency in the normal state.  
The temperature goes back to the initial temperature of the $^3$He-$^4$He mixture within several tens of ms, since the resonance frequency of the O(1)$_{\parallel}$ NMR peak was shifted to the frequency observed at 70 mK.
The relaxation time of the electronic spin temperature after heating is nearly the same in the RF-pulse energy range from 50 to150 $\mu$J, as shown in Fig. \ref{figA2} (b).

Now we discuss the temperature recovery process after the heat pulse on the basis of the previous discussion on the Pt-NMR below 1 K \cite{Walstedt1964}.
The difficulty of the low-temperature NMR measurements on metallic and superconducting compounds mainly arises from the eddy current heating, which affects the low-temperature results seriously as in the present case.
Let us consider that the single crystalline Sr$_2$RuO$_4$ immersed in the $^3$He-$^4$He mixture is in equilibrium at 70 mK, and what happens after the application of an RF pulse based on the above experimental results.
First, during the RF-pulse applicaton, (1) dissipation of the RF-pulse energy causes the temperature of the surface where the RF field penetrates to rise instantaneously although the temperature of the nuclear spin remains nearly at the initial temperature since the $T_1$ is several orders of magnitude longer than this time scale. 
Next, (2) the temperature is immediately homogenized within the Sr$_2$RuO$_4$ sample and the heat is transferred from Sr$_2$RuO$_4$ to the $^3$He-$^4$He mixture. 
Thus, the temperature of Sr$_2$RuO$_4$ comes back to nearly the initial temperature.
 
When considering the first process, we need to know how much heat is absorbed in Sr$_2$RuO$_4$ by the RF pulse through the NMR coil.
However, it is a quite complex process and difficult to quantify this amount of energy since various factors such as the sample shape and the surface roughness are involved.
Nevertheless, it is possible to estimate the necessary energy (injected heat $Q$) to raise the temperature of Sr$_2$RuO$_4$ beyond $T_{c}$($H$) $\sim 1.2$ K.
Assuming that Sr$_2$RuO$_4$ is thermally isolated during the immediate heat-up by the RF-pulse, we estimate the injected heat $Q$, which raises the initial temperature $T_0$ to the highest temperature $T'$:
\begin{equation*}
Q=\int_{T_0}^{ T'}{C(T) dT}.
\end{equation*}
At the initial temperature $T_0$ = 70 mK, the electronic heat capacity of Sr$_2$RuO$_4$ is small because it is deep in the SC state, and we assume that the injection of $Q$ immediately destroys the superconductivity in the whole region of the sample. 
Because of the specific-heat jump at $T_c$, we use the relation $C(T) = N~\gamma_{\rm el}~T$ for the whole temperature region, where $\gamma_{\rm el}$ is the molar specific heat constant in the normal state of Sr$_2$RuO$_4$ and $N_{\rm mol}$ is the number of moles in the sample.
Thus,
\begin{equation*}
Q=\frac{N_{\rm mol} \gamma_\text{el}}{2}\left(T'^2-T_0^2\right).
\end{equation*}
Using the following parameters, $\gamma_e = 38$ mJ / mol K$^2$, $N = 2.6 \times 10^{-4}$ mol and $T' = T_c (0.7 T)~{\sim 1.2}$ K , $Q$ is estimated as 7.1 $\mu$ J.
It is noteworthy that the estimated $Q$ is smaller than the threshold $E_{\rm thre} \sim 40~\mu$J, below which the Knight shift starts to decrease as seen in Fig. \ref{fig1}. 
From this estimation, it is roughly evaluated that effectively $\sim 18$\% of RF-pulse energy is absorbed in the sample to raise its temperature. 

Next, we consider the process (2) quantitatively. 
We assume that heat delivered to Sr$_2$RuO$_4$ by the RF pulse is immediately diffused into the entire sample and then released from its surface to the thermal reservoir of the $^3$He-$^4$He mixture. 
In this case, it is known that the Kapitza resistance is operating at the interface between $^4$He and the metal, and the rate of heat transfer $\kappa$ is given by 
\begin{equation*}
\kappa(T, T_0) = \alpha A (T^3-T_0^3),
\end{equation*}
where $A$ is the surface area of Sr$_2$RuO$_4$, and $\alpha$ is a constant related to an inverse of the Kapitza resistance. 
The differential equation for $T(t)$ is 
\begin{eqnarray*}
-C(T) dT& =& \kappa(T(t), T_0) dt \\
-N_\text{mol}\gamma_\text{el} T(t) dT & = &\alpha A [T(t)^3-T_0^3] dt
\end{eqnarray*}
With the expressions given for $C$($T$) and $\kappa(T(t), T_0)$, the relation between $t$ and $T(t)$ is expressed as follows\cite{Walstedt1964}:
\begin{eqnarray}
\frac{3 \alpha A T_0 t}{\gamma_\text{el} N_\text{mol}}\int_{0}^{t}dt & = & -\int_{T(0)}^{T(t)}\frac{3 T_0 T(t) dT}{T(t)^3-T_0^3}  \nonumber \\
\frac{3 \alpha A T_0 t}{\gamma_{\rm el}~N_{\rm mol}} & = & -\ln{\left[~\frac{T(t)-T_0}{T(0)-T_0}\right]} +\frac{1}{2}\ln{\left[~\frac{T(t)^2+T(t)T_0+T_0^2}{T(0)^2+T(0)T_0+T_0^2}\right]} \\
 &  & -\sqrt{3}~\left[\tan^{-1}\left(\frac{2T(t)/T_0+1}{\sqrt{3}}\right)-\tan^{-1}\left(\frac{2T(0)/T_0+1}{\sqrt{3}}\right)\right] \nonumber.
\end{eqnarray}
The relaxation time $\tau_{\rm theo}$ of the temperature recovery back to the initial value is expressed with 
\begin{equation*}
\tau_{\rm theo} = \frac{N_{\rm mol} \gamma_{\rm el}}{ 3\alpha A T_0}.
\end{equation*}

Using the following parameters\cite{LittlePR1961} $\alpha =  2 \times 10^{4}$ erg cm$^{-2}$ s$^{-1}$ K$^{-3}$ and $A = 0.73$ cm$^2$, we obtain $\tau_{\rm theo} = 32.5$ ms.
After the heat-up pulse, we assume that the temperature immediately increases to highest $T'$ during a negligibly short time, and thus $T(0) = T'$, and goes back to $T_c \sim 1.2$ K at a time when the resonance frequency starts to shift. 
It is noted that until this time is reached, the resonance frequency is constant due to the $T$-independent Knight shift in the normal state.       
If we adopt that a 10\% RF-pulse energy is used to heat the sample, $T'$ would be estimated to be 1.74, 1.42, and 1.00 K for the RF-pulse energy of 150, 100 and 50 $\mu$J, respectively.
The relaxation of temperature [$T(t)$] is estimated by numerically solving the eq. (B$\cdot$1) and the results are shown with the dotted curves in Fig.~\ref{figA2} (b). 
Fairy good simulation curves are obtained by this simplified model.
We can roughly understand the temperature increase and recovery to the initial temperature after the large-energy RF pulse, quantitatively.                      
        
%%%%%%%%%%%%%%%%%%%%%%%%%%%%%%%%%%%%%%%%%%%% Appendix2 %%%%%%%%%%%%%%%%%%%%%%%%%%%%
%%%%%%%%%%%%%%%%%%%%%%%%%%%%%%%%%%%%%%%%%%%%%%%%%%%%%%%%%%%%%%%%%%%%%%%%%%%%%%%

\section{Eddy-current heating in the normal and SC states of Sr$_2$RuO$_4$}
In Appendix D, we estimate the eddy-current heating in the normal and SC state of Sr$_2$RuO$_4$.
By applying RF pulses, the metallic sample in the solenoid coil is instantaneously heated up by the eddy current. 
The measured sample has a layered shape with highly anisotropic resistivity, and thus the power $P_{\rm v}$ per a unit volume can be approximately expressed as 
\begin{equation*}
P_v =\frac{1}{2}H_1^2\mu f \left[\frac{\delta_c}{2t_c} \frac{\sinh{(t_c/\delta_c)}-\sin{(t_c/\delta_c)}}{\cosh{(t_c/\delta_c)}+\cos{(t_c/\delta_c)}}+\frac{\delta_a}{2t_a} \frac{\sinh{(t_a/\delta_a)}-\sin{(t_a/\delta_a)}}{\cosh{(t_a/\delta_a)}+\cos{(t_a/\delta_a)}}\right] \times 10^{-5}   \hspace{0.5cm}   \mbox{  W/cm$^3$ }
\end{equation*}      
with the skin depth $\delta_{c, (a)}$ along the $c, (a)$-axis direction of
\begin{equation*}
\delta_{c, (a)} = \sqrt{\frac{10^2 \rho_{ab, (c)}}{4 \pi^2 \times 10^{-7}f}} \hspace{1cm} \mbox{cm}.
\end{equation*}
Here, $H_1$ in mT is an alternative magnetic field in the solenoid coil, $\mu$ is relative magnetic permeability of $1+ \chi \sim 1$, $f$ in Hz is the frequency of the alternative field (NMR frequency), 
$t_{a, (c)}$ in cm is a length along the $a$ ($c$) axis direction of the plate-shaped sample, $\rho_{ab, (c)}$ in $\Omega \cdot$ cm is the resistivity along the $ab$ plane and along the $c$ axis, respectively.
It is noted that most eddy current heating in Sr$_2$RuO$_4$ is induced by the conduction along the $c$ axis, which is usually ignored in a layered sample with isotropic resistivity.   
When the pulse width $t_w$ of the $\pi/2$ pulse is 7 $\mu$s, $H_1$ is estimated to be 6.2 mT from the relation of $\pi /2 = \gamma_N H_1 t_w$.
In this case, the total heat $Q = P_v~V~t_w$ of the sample volume $V$ is estimated to be 15 $\mu$J, where $\rho_{ab, (c)}$ = 0.1 (1500) $\mu~\Omega \cdot$ cm is used. 
When $T_0$ before the RF pulse is 1.5 K, this heat makes $T'$ of the electrons system just after the RF pulse increase up to 2.27 K if all heat is delivered to the sample.   
However, since the sample is immersed in a superfluid $^4$He, the heating-effect would be somewhat suppressed, and the temperature decreases immediately back to $T_0$.

\begin{figure}[!tbh]
\begin{center}
\includegraphics[width=8.5cm,clip]{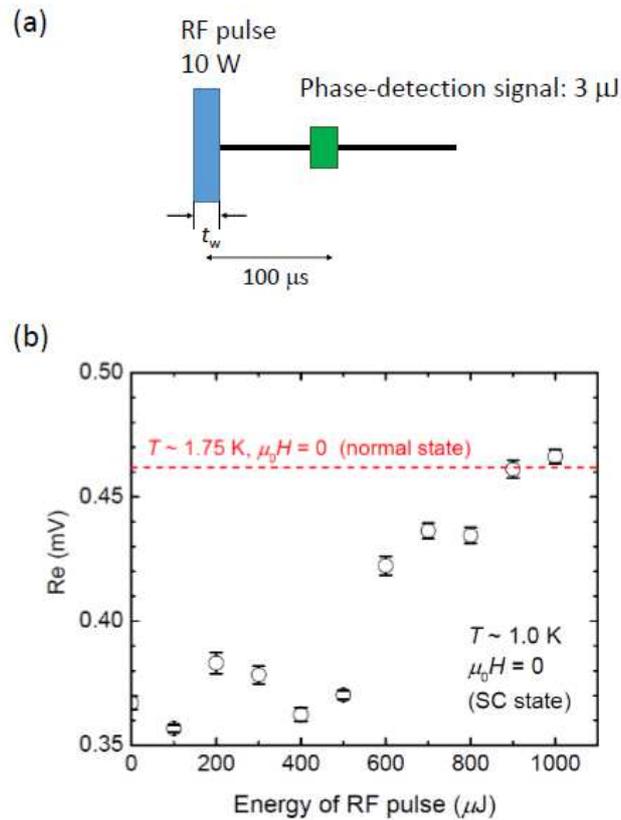}
\end{center}
\caption{(Color online) (a) The pulse sequence for the heat-up test in the SC state (1 K). The nominal power of the RF pulse was fixed to be 10 W (-10 dB) and the energy of the RF pulse was changed by changing the pulse interval $t_w$. To detect the destruction of superconductivity by the RF pulse, the phase-detection RF signal with 3 $\mu$J was applied at 100 $\mu$s after the RF pulse and the phase of the NMR tank circuit was measured with the NMR receiver. 
(b)Dependence of the real (in-phase) part of the phase-detection RF signal on the energy of the RF pulse. The superconductivity is not destroyed up to 900 $\mu$J. This is nominal value and indicates that the RF-pulse energy cannot be delivered to the sample due to the Meissner shielding.       }
\label{figA3}
\end{figure}

When the sample becomes SC, the RF heating effect to the whole sample becomes much weaker, because the eddy current disappears in the SC state, and the RF field can penetrate within the region of SC penetration-length $\lambda$, much shorter than $\delta_{ab, (c)}$.
Actually when the heat-up effect by a RF pulse was examined at 1 K in the SC state by the pulse sequence showing in Fig.~\ref{figA3}, the Meissner signal from the sample can be observed up to much larger RF-pulse energy.
This is because the penetration of the RF field is limited within $\lambda$, and the heat is diffused into the SC part of the sample.
However, this does not necessarily mean that the heat-up effect of the NMR measurements in the SC state becomes weak, since NMR probes the nuclear spins in the region of $\lambda$, which are affected by the RF-pulse power.
It is noted that the superconductivity is suppressed by a much smaller RF-pulse power when the sample is in the magnetic field as shown in Fig.~\ref{figA1}.
When the fields are applied exactly parallel to the $ab$ plane in Sr$_2$RuO$_4$, the vortices penetrate in the ``lock-in '' state due to the 2-D character of the electronic properties, and the Meissner shielding becomes very weak, as shown in Fig. \ref{fig1} (b).
In addition, the imaginary part of the ac susceptibility is large in the ``lock-in'' state\cite{YoshidaJPSJ1996}.
Thus, this would enhance the heat-up effect by the RF-pulse significantly when $H$ is applied to the $ab$ plane in the SC state.

%%%%%%%%%%%%%%%%%%%%%%%%%%%%%%%%%%%%%%%%%%%% Appendix3 %%%%%%%%%%%%%%%%%%%%%%%%%%%%
%%%%%%%%%%%%%%%%%%%%%%%%%%%%%%%%%%%%%%%%%%%%%%%%%%%%%%%%%%%%%%%%%%%%%%%%%%%%%%%                  

%\bibliography{RefSRO}

\end{document}